\documentclass[12pt]{article}
\usepackage{amssymb}
\usepackage{amsmath}
\def\d{\delta}

\def\be{\begin{equation}}
\def\ee{\end{equation}}
\def\arr{\begin{array}{rll}}
\def\ea{\end{array}}
\def\bea{\begin{eqnarray}}
\def\eea{\end{eqnarray}}

\def\N2{$N{=}2$}

\def\>{\rangle}
\def\<{\langle}
\def\+{\dagger}
\def\={\ =\ }

\begin{document}
\renewcommand{\thefootnote}{\arabic{footnote}}
\begin{titlepage}
\setcounter{page}{0}
\begin{flushright}
$\qquad$
\end{flushright}
\vskip 3cm
\begin{center}
{\LARGE\bf{An alternative Hamiltonian formulation }}\\
\vskip 0.7cm
{\LARGE\bf{for the Pais-Uhlenbeck oscillator}}\\
\vskip 1cm
$
\textrm{\Large Ivan Masterov\ }
$
\vskip 0.7cm
{\it
Laboratory of Mathematical Physics, Tomsk Polytechnic University, \\
634050 Tomsk, Lenin Ave. 30, Russian Federation}
\vskip 0.7cm
{E-mail: masterov@tpu.ru}

\end{center}
\vskip 1cm
\begin{abstract}
\noindent
Ostrogradsky's method allows one to construct Hamiltonian formulation for a higher derivative system. An application of this approach to the Pais-Uhlenbeck oscillator yields the Hamiltonian which is unbounded from below. This leads to the ghost problem in quantum theory. In order to avoid this nasty feature, the technique previously developed in [Acta Phys. Polon. B 36 (2005) 2115] is used to
construct an alternative Hamiltonian formulation for the multidimensional Pais-Uhlenbeck oscillator of arbitrary even order with distinct frequencies of oscillation. This construction is also generalized to the case of an $\,\mathcal{N}=2$ supersymmetric Pais-Uhlenbeck oscillator.
\end{abstract}


\vskip 0.5cm


\end{titlepage}

\noindent
{\bf{\large 1. Introduction}}
\vskip 0.5cm

Higher derivative theories attract interest mostly due to their nice renormalization properties \cite{Thiring,Stelle}.
The method to construct Hamiltonian formulation for such systems has been proposed by Ostrogradsky \cite{Ostrogradski}.
In general, Hamiltonians obtained in such a way contain terms linear in momenta and are unbounded from below.
This leads to the ghost problem on quantization \cite{Pais,Woodard}.
The desire to cure this problem stimulates the investigation of the Pais-Uhlenbeck (PU) oscillator \cite{Pais}.

After applying an appropriate canonical transformation \cite{Pais,Smilga}, Ostrogradsky's Hamiltonian for the multidimensional PU oscillator of order $2n$ with distinct frequencies of oscillation $\omega_k$, $k=0,1,..,n-1$, takes the form
\bea\label{mh}
H=\frac{1}{2}\sum_{k=0}^{n-1}(-1)^{k+1}(p_i^k p_i^k+\omega_k^2 x_i^k x_i^k).
\eea
When conventional quantization scheme is applied, the harmonic oscillators with negative overall factor bring about troubles with unbounded from below energy spectrum and, hence, with the absence of the ground state  \cite{Pais,Woodard}.
This motivates a search for an alternative Hamiltonian formulation and quantization procedure which lead to physically viable quantum theory \cite{Kosinski}-\cite{Reyes}.
So far the efforts have been focused mostly on the one-dimensional PU oscillator of the fourth order \cite{Kosinski}-\cite{KLS}, \cite{Reyes}.
In particular, an elegant method to obtain an alternative canonical formalism with positive-definite Hamiltonian has been formulated in \cite{Kosinski}.
This alternative formulation has been realized in two steps.
At the first stage, two functionally independent integrals of motion which are quadratic in variables have been used so as to write down an ansatz for the Hamiltonian of the fourth-order PU oscillator.
The second step implies the derivation of an appropriate Poisson structure.

An attempt to generalize the results in \cite{Kosinski} to the case of higher order PU oscillator has been made in \cite{Damaskinsky}. However, this generalization exhibits some features which seem to contradict each other. On the one hand, the alternative Hamiltonian in \cite{Damaskinsky} is not positive definite. In this sense it is not better than Ostrogradsky's one. On the other hand, it was claimed in \cite{Damaskinsky} that the quantum theory of the PU oscillator constructed with the use of the alternative Hamiltonian is ghost free. As will be demonstrated below, the reason is that the alternative Hamiltonian together with the Poisson structure in \cite{Damaskinsky} do not reproduce the equation of motion of the original PU oscillator.

One of the goals of the present paper is to use the technique introduced in \cite{Kosinski} is order to obtain an alternative Hamiltonian formulation for the multidimensional PU oscillator of arbitrary even order with distinct frequencies of oscillation. We also explain which claims in \cite{Damaskinsky} are incorrect.

Recently, in  \cite{Masterov}, \cite{Masterov_1} an $\,\mathcal{N}=2$ supersymmetric extension of the PU oscillator has been constructed.
It has been shown that the invariance of the model under time translations implies unbounded from below spectrum.
The Hamiltonian can be presented as the sum of decoupled $\,\mathcal{N}=2$ supersymmetric harmonic oscillators with alternating sign.
The corresponding quantum theory is characterized by the presence of negative-norm states and by the absence of the ground state.
Our second concern in this paper is the construction of an alternative Hamiltonian for an $\,\mathcal{N}=2$ supersymmetric PU oscillator which is achieved by generalizing
the method in \cite{Kosinski} to the $\,\mathcal{N}=2$ supersymmetric case.

The paper is organized as follows.
In the next section we apply the method previously developed in Ref. \cite{Kosinski} to obtain an alternative Hamiltonian formulation for the PU oscillator of order $2n$.
In Section 3, in the same manner we construct an alternative Hamiltonian formalism for an $\,\mathcal{N}=2$ supersymmetric PU oscillator.
We summarize our results and discuss possible further developments in the concluding Section 4.
Some technical details are given in Appendix.

\vskip 0.5cm
\noindent
{\bf{\large 2. An alternative Hamiltonian formulation for the PU oscillator}}
\vskip 0.5cm

The equation of motion of the multidimensional PU oscillator of order $2n$ can be written in the following form \cite{Pais}
\bea\label{EOM}
\prod_{k=0}^{n-1}\left(\frac{d^2}{dt^2}+\omega_k^2\right)x_i=\sum_{k=0}^{n}\sigma_k^n x_i^{(2k)}=0,\quad\mbox{where}\quad
\sigma_k^n=\sum_{i_1<i_2<..<i_{n-k}}\omega_{i_1}^2\omega_{i_2}^2..\omega_{i_{n-k}}^2,\quad\sigma_{n}^{n}=1,
\eea
where $i=1,2,..,d$ is a spatial index, while the index in braces denotes the order of time derivative. For definiteness, we choose $0<\omega_0<\omega_1<..<\omega_{n-1}$. This equation can be obtained from the action functional\footnote{Throughout the work the summation over repeated spatial indices is understood.}
\bea\label{act}
S=\frac{1}{2}\int dt\; x_i\prod_{k=0}^{n-1}\left(\frac{d^2}{dt^2}+\omega_k^2\right)x_i,
\eea
which is invariant under time translations. The Noether theorem yields the integral of motion \cite{Pais}
\bea\label{H}
H=\sum_{k=0}^{n-1}(-1)^{k+1}J_k,
\eea
where we denoted
\bea\label{J}
J_k=\frac{\rho_k}{2}\left(\prod_{m=0\atop m\neq k}^{n-1}\left(\frac{d^2}{dt^2}+\omega_m^2\right)\frac{d x_i}{dt}\right)^{2}+\frac{\rho_k\omega_k^2}{2}\left(\prod_{m=0\atop m\neq k}^{n-1}\left(\frac{d^2}{dt^2}+\omega_m^2\right)x_i\right)^2,\; \rho_k=\frac{(-1)^k}{\prod\limits_{m=0\atop m\neq k}^{n-1}(\omega_{m}^{2}-\omega_{k}^{2})}.
\eea
It is easy to see that all $\rho_k$ are positive. The quantities $J_k$ are positive-definite integrals of motion which correspond to the symmetry transformations
\bea
\d x_i=-\rho_k\prod_{m=0\atop m\neq k}^{n-1}\left(\frac{d^2}{dt^2}+\omega_m^{2}\right)\frac{d x_i}{dt}\,\epsilon_k,
\nonumber
\eea
where $\epsilon_k$, $k=0,1,..,n-1$, are infinitesimal parameters. The Noether integral of motion which corresponds to the invariance under time translations is unbounded from below.
This implies a similar property for Ostrogradsky's Hamiltonian which is the phase space analogue of (\ref{H}). In quantum theory one reveals the well known problems which were discussed in the Introduction.

In the next subsection we remind some basic facts about the approach in \cite{Kosinski} which leads to an alternative Hamiltonian formulation for the fourth-order PU oscillator.

\vskip 0.5cm
\noindent
{\bf{2.1. The fourth-order PU oscillator}}
\vskip 0.5cm

According to \cite{Kosinski}, one can use the quadratic integrals of motion $J_0$ and $J_1$ so as to write down an ansatz for an alternative Hamiltonian
\bea\label{Ans}
\mathcal{H}_2=\alpha_0 J_0+\alpha_1 J_1,
\eea
where $\alpha_0$ and $\alpha_1$ are arbitrary nonzero constants. This constant of the motion can play the role of the Hamiltonian provided the relations
\bea\label{1111}
\{x_i^{(k)},\mathcal{H}_2\}=x_i^{(k+1)},\quad\mbox{for}\quad k=0,1,2;\qquad \{x_i^{(3)},\mathcal{H}_2\}=-(\omega_0^2+\omega_1^2)x_i^{(2)}-\omega_0^2\omega_1^2 x_i
\eea
hold with respect to some Poisson bracket $\{\cdot,\cdot\}$. The latter is to be determined \cite{Kosinski}. The conditions above can be expressed in the form of a system of linear algebraic equations. A unique solution to it yields the following nonvanishing Poisson--like structure relations for the variables of the configuration space \cite{Kosinski}
\bea\label{PB4}
\begin{aligned}
&
\{x_i,x_j^{(1)}\}=\frac{1}{\omega_1^2-\omega_0^2}\left(\frac{1}{\alpha_0}+\frac{1}{\alpha_1}\right)\d_{ij},&& \{x_i,x_j^{(3)}\}=-\frac{1}{\omega_1^2-\omega_0^2}\left(\frac{\omega_0^2}{\alpha_0}+\frac{\omega_1^2}{\alpha_1}\right)\d_{ij},
\\[5pt]
&
\{x_i^{(1)},x_j^{(2)}\}=\frac{1}{\omega_1^2-\omega_0^2}\left(\frac{\omega_0^2}{\alpha_0}+\frac{\omega_1^2}{\alpha_1}\right)\d_{ij},&&
\{x_i^{(2)},x_j^{(3)}\}=\frac{1}{\omega_1^2-\omega_0^2}\left(\frac{\omega_0^4}{\alpha_0}+\frac{\omega_1^4}{\alpha_1}\right)\d_{ij}.
\end{aligned}
\eea
Evidently, this Poisson--like structure together with (\ref{Ans}) correspond to Ostrogradsky's Hamiltonian formalism if $\alpha_0=-1$, $\alpha_1=1$.

The canonical coordinates with respect to the Poisson structure (\ref{PB4}) have the form \cite{Pais}
\bea\label{var4}
\begin{aligned}
&
x_i^0=\sqrt{\frac{|\alpha_0|}{\omega_1^2-\omega_0^2}}\left(x_i^{(2)}+\omega_1^2 x_i\right),\qquad p_i^0=sign(\alpha_0)\frac{d x_i^0}{dt};
\\[5pt]
&
x_i^1=\sqrt{\frac{|\alpha_1|}{\omega_1^2-\omega_0^2}}\left(x_i^{(2)}+\omega_0^2 x_i\right),\qquad p_i^1=sign(\alpha_1)\frac{d x_i^1}{dt},
\end{aligned}
\eea
where $sign(x)$ is the standard signum function.
Indeed, it is straightforward to verify that these variables obey the following nonvanishing structure relations
\bea\label{canon}
\{x_i^k,p_j^m\}=\d_{ij}\d_{km}
\eea
under the Poisson bracket (\ref{PB4}).

The Hamiltonian (\ref{Ans}) in terms of the canonical variables (\ref{var4}) takes the form
\bea
\mathcal{H}_2=\frac{1}{2}sign(\alpha_0)(p_i^0 p_i^0+\omega_0^2 x_i^0 x_i^0)+\frac{1}{2}sign(\alpha_1)(p_i^1 p_i^1+\omega_1^2 x_i^1 x_i^1).
\nonumber
\eea
If $\alpha_0$ and $\alpha_1$ are both positive, then one has a positive-definite Hamiltonian. This obviously leads to the ghost-free quantum theory for the fourth-order PU oscillator.

\vskip 0.5cm
\noindent
{\bf{2.2. The general case}}
\vskip 0.5cm

In order to generalize an alternative Hamiltonian structure obtained in \cite{Kosinski} to the case of the PU oscillator of order $2n$, let us introduce the following integral of motion
\bea\label{HH}
\mathcal{H}_n=\sum_{k=0}^{n-1}\alpha_k J_k,
\eea
where constants $\alpha_k$ with $k=0,1,..,n-1$ can take arbitrary nonzero values.

At the next step, we have to search for such a Poisson structure which produces the equations
\bea\label{canon2}
\{x_i^{(k)},\mathcal{H}_n\}=x_i^{(k+1)},\quad k=0,1,..,2n-2,\qquad \{x_i^{(2n-1)},\mathcal{H}_n\}=-\sum_{k=0}^{n-1}\sigma_{k}^{n}x_i^{(2k)}.
\eea
In contrast to the analysis in the previous subsections, a straightforward calculation of this structure faces technical difficulties.
Note, however, that (\ref{PB4}) can be written in a more compact form
\bea\label{PB}
\{x_i^{(s)},x_j^{(m)}\}=w_{sm}\d_{ij}=\left\{\begin{aligned}
&
0,&& \mbox{$s+m$ -- even};\\[5pt]
&
(-1)^\frac{s+m-(-1)^{s}}{2}\sum_{k=0}^{n-1}\frac{\omega_k^{s+m-1}\rho_k}{\alpha_k}
\d_{ij},&& \mbox{$s+m$ -- odd}.
\end{aligned}\right.
\eea
The corresponding Poisson bracket is defined in the standard way
\bea\label{PS}
\{A,B\}=\sum_{s,m=0}^{2n-1}w_{sm}\frac{\partial A}{\partial x_i^{(s)}}\frac{\partial B}{\partial x_i^{(m)}}.
\eea
One can straightforwardly verify that the equations (\ref{canon2}) are satisfied with respect to this bracket by making use of the following identities\footnote{The proof of the identities (\ref{i}) is given in Appendix.}
\bea\label{i}
\begin{aligned}
&
\sum_{m=0}^{n-1}(-1)^m\omega_p^{2m}\sigma_{m,k}^{n}=\frac{(-1)^k}{\rho_k}\d_{kp};
\\[5pt]
&
\sum_{k=0}^{n-1}(-1)^k(-\omega_k^2)^{s}\sigma_{p,k}^{n}\rho_k=
\left\{
\begin{aligned}
&
\d_{sp},&& s=0,1,..,n-1;
\\[5pt]
&
-\sigma_p^n,&& s=n,
\end{aligned}
\right.
\end{aligned}
\eea
where we denote
\bea
\sigma_{p,k}^n=\sum_{i_1<i_2<..<i_{n-p-1}=0\atop i_1,i_2,..,i_{n-p-1}\neq k}^{n-1}\omega_{i_1}^2\omega_{i_2}^2..\omega_{i_{n-p-1}}^2,\qquad \sigma_{n-1,k}^n=1.
\nonumber
\eea

One can also show with the aid of these identities that the variables
\bea\label{var}
x_i^{k}=\sqrt{|\alpha_k|\rho_k}\prod_{m=0\atop m\neq k}^{n-1}\left(\frac{d^2}{dt^2}+\omega_m^{2}\right)x_i,\quad p_i^k=sign(\alpha_k)\frac{d x_i^k}{dt},\qquad k=0,1,..,n-1,
\eea
obey the structure relations (\ref{canon}) under the Poisson bracket (\ref{PS}). The existence of these coordinates implies also that the Jacobi identity is satisfied for the structure (\ref{PB}).

The Hamiltonian (\ref{HH}) can be rewritten in the following form
\bea\label{Hcanon}
\mathcal{H}_n=\frac{1}{2}\sum_{k=0}^{n-1}sign(\alpha_k)(p_i^k p_i^k+\omega_k^2 x_i^k x_i^k).
\eea
The choice $\alpha_k=(-1)^{k+1}$ corresponds to Ostrogradsky's approach. But if all constants are positive, the alternative Hamiltonian is positive-definite and is more suitable for physical applications.

In Refs. \cite{KLS,KL}, in order to obtain an alternative Hamiltonian formulation of the PU oscillator, another approach is used which is based on the observation that the equation of motion (\ref{EOM}) is equivalent to a system of second-order differential equations which describe a set of decoupled harmonic oscillators. Then the Hamiltonian formulation of the latter system is linked to the PU oscillator as well. The Hamiltonian (\ref{Hcanon}) coincides with that in \cite{KLS,KL}.

\vskip 0.5cm
\noindent
{\bf{2.3. About the results in Ref. \cite{Damaskinsky}}}
\vskip 0.5cm

Let us consider an alternative Hamiltonian structure for the one-dimensional PU oscillator of the order $2n$ which has been obtained in Ref. \cite{Damaskinsky}. The authors introduce the alternative Hamiltonian of the form
\bea
\tilde{H}=\sum_{k=0}^{n-1}b_k\tilde{H}_k,
\nonumber
\eea
where
\bea\label{bk}
b_k=\frac{1}{\omega_k\prod\limits_{j=0\atop j\neq k}^{n-1}(\omega_k^2-\omega_j^2)},\qquad \tilde{H}_k=\frac{1}{2}\left(\sum_{m=0}^{n-1}\sigma_{m,k}^{n}x^{(2m+1)}\right)^2+\frac{\omega_k^2}{2}\left(\sum_{m=0}^{n-1}\sigma_{m,k}^{n}x^{(2m)}\right)^2.
\eea
It is evident that $\tilde{H}$ is not positive definite because for each value of $n=2,3,..$ there are negative coefficients among $b_k$. In particular, for the fourth-order case one finds
\bea\label{HDS}
\tilde{H}=\frac{1}{\omega_0(\omega_0^2-\omega_1^2)}\tilde{H}_0+\frac{1}{\omega_1(\omega_1^2-\omega_0^2)}\tilde{H}_1
=\frac{1}{\omega_0\omega_1(\omega_1^2-\omega_0^2)}(\omega_1\tilde{H}_0-\omega_0\tilde{H}_1).
\eea
In this sense the Hamiltonian $\tilde{H}$ is not better than Ostrogradsky's one (\ref{H}). It comes as a surprise that the quantum analogue of (\ref{HDS}) presented in \cite{Damaskinsky} has the form (Eq. (40) in \cite{Damaskinsky})
\bea\label{qth}
\hat{H}=\hbar\sum_{k=0}^{1}\omega_k\left(a_k^{\dag}a_k+\frac{1}{2}\right),
\eea
where $a_k$, $a_k^{\dag}$ are the creation and annihilation operators which obey the conventional commutation relations. It is obvious that the quantum theory determined by (\ref{qth}) is ghost-free.

It was claimed in \cite{Damaskinsky} that the Poisson structure
\bea\label{PBB}
\begin{aligned}
&
\{x,x^{(1)}\}=2(\omega_0+\omega_1), && \{x,x^{(3)}\}=-2(\omega_0^3+\omega_1^3),
\\[3pt]
&
\{x^{(1)},x^{(2)}\}=2(\omega_0^3+\omega_1^3), && \{x^{(2)},x^{(3)}\}=2(\omega_0^5+\omega_1^5)
\end{aligned}
\eea
yield equations (\ref{1111}) which involve the Hamiltonian (\ref{HDS}). However, the straightforward calculations give
\bea
\begin{aligned}
&
\{x,\tilde{H}\}=-2(\omega_0^2+\omega_1^2)x^{(1)}-4x^{(3)}, && \{x^{(1)},\tilde{H}\}=4\omega_0^2\omega_1^2 x+2(\omega_0^2+\omega_1^2)x^{(2)},
\\[3pt]
&
\{x^{(2)},\tilde{H}\}=4\omega_0^2\omega_1^2 x^{(1)}+2(\omega_0^2+\omega_1^2)x^{(3)}, && \{x^{(3)},\tilde{H}\}=-2\omega_0^2\omega_1^2(\omega_0^2+\omega_1^2)x-2(\omega_0^4+\omega_1^4)x^{(2)},
\end{aligned}
\nonumber
\eea
which differ from (\ref{1111}).

Suppose the coefficients $b_k$ are allowed to be amended, while the Poisson structure is fixed in the form (\ref{PB}). With the aid of (\ref{PB4}) one can find the Hamiltonian which corresponds to the structure (\ref{PBB}). Demanding  the right hand sides in Eqs. (\ref{PB4}) and (\ref{PBB}) to be equal to each other, one gets the system of three linear equations involving two variables $\alpha_0$ and $\alpha_1$
\bea
&&
\frac{1}{\omega_1^2-\omega_0^2}\left(\frac{1}{\alpha_0}+\frac{1}{\alpha_1}\right)=2(\omega_0+\omega_1), \qquad
\frac{1}{\omega_1^2-\omega_0^2}\left(\frac{\omega_0^2}{\alpha_0}+\frac{\omega_1^2}{\alpha_1}\right)=2(\omega_0^3+\omega_1^3),
\nonumber
\\[3pt]
&&
\qquad\qquad\qquad\qquad\quad\frac{1}{\omega_1^2-\omega_0^2}\left(\frac{\omega_0^4}{\alpha_0}+\frac{\omega_1^4}{\alpha_1}\right)=2(\omega_0^5+\omega_1^5),
\nonumber
\eea
which has the solution
\bea
\alpha_k=\frac{1}{2\omega_k(\omega_1^2-\omega_0^2)}.
\nonumber
\eea
Taking into account (\ref{J}), one finally concludes that the correct choice of the coefficients $b_k$ in (\ref{bk}) for the fourth-order case is
\bea
b_k=\frac{1}{2\omega_k(\omega_1^2-\omega_0^2)^2}.
\nonumber
\eea
Higher-order PU oscillators can be treated likewise.

\newpage

\noindent
{\bf{2.4. Compatible generalizations}}
\vskip 0.5cm

Consider the deformation of the Hamiltonian (\ref{HH})
\bea\label{HHHH}
\mathcal{H}_{int}=\mathcal{H}_n+U,
\eea
where $U=U(x_i,x_i^{(1)},..,x_i^{(2n-1)})$ is an arbitrary function. Let us require this function to obey the equations
\bea
\{x_i^{(k)},U\}=0,\quad k=0,1,..,2n-2
\nonumber
\eea
under the bracket (\ref{PS}). These conditions can be presented in the form
\bea
&&\label{odd}
\sum_{m=0}^{n-1}(-1)^{m}\frac{\partial U}{\partial x_i^{(2m+1)}}\sum_{k=0}^{n-1}\frac{\omega_k^{2s+2m}\rho_k}{\alpha_k}=0,\qquad s=0,1,..,n-1,
\\[5pt]
&&\label{even}
\sum_{m=0}^{n-1}(-1)^{m}\frac{\partial U}{\partial x_i^{(2m)}}\sum_{k=0}^{n-1}\frac{\omega_k^{2s+2m}\rho_k}{\alpha_k}=0,\qquad\quad s=0,1,..,n-2.
\eea
Subsystems (\ref{odd}) and (\ref{even}) can be treated separately. The first subsystem (\ref{odd}) is homogeneous and includes $n$ linear equations involving $n$ partial derivatives of the function $U$. The matrix of this subsystem can be represented as follows
\bea
A=\left(
\begin{aligned}
& 1 && 1 && ... && 1
\\[5pt]
& \omega_0^2 && \omega_1^2 && ... && \omega_{n-1}^2
\\[5pt]
& ... && ... && ... && ...
\\[5pt]
& \omega_0^{2(n-1)} && \omega_1^{2(n-1)} && ... && \omega_{n-1}^{2(n-1)}
\end{aligned}
\right)
\left(
\begin{aligned}
& \frac{\rho_0}{\alpha_0} && 0 && ... && 0
\\[5pt]
& 0 && \frac{\rho_1}{\alpha_1} && ... && 0
\\[5pt]
& ... && ... && ... && ...
\\[5pt]
& 0 && 0 && ... && \frac{\rho_{n-1}}{\alpha_{n-1}}
\end{aligned}
\right)
\left(
\begin{aligned}
& 1 && -\omega_0^2 && ... && (-\omega_{0}^2)^{n-1}
\\[5pt]
& 1 && -\omega_{1}^2 && ... && (-\omega_{1}^2)^{n-1}
\\[5pt]
& ... && ... && ... && ...
\\[5pt]
& 1 && -\omega_{n-1}^2 && ... && (-\omega_{n-1}^2)^{n-1}
\end{aligned}
\right).
\nonumber
\eea
The corresponding determinant reads
\bea
det\,A=(-1)^{\frac{n(n-1)}{2}}\left(\prod_{i_1<i_2=0}^{n-1}(\omega_{i_2}^2-\omega_{i_1}^{2})\right)^2\prod\limits_{i=0}^{n-1}\frac{\rho_i}{\alpha_i}=
\frac{(-1)^{\frac{n(n-1)}{2}}}{\prod\limits_{i=0}^{n-1}\alpha_i}.
\nonumber
\eea
Then the matrix of the subsystem (\ref{odd}) is nondegenerate and one has only the trivial solution
\bea\label{sol}
\frac{\partial U}{\partial x_i^{(2m+1)}}=0,\; \mbox{with}\; m=0,1,..,n-1.
\eea

The subsystem (\ref{even}) is a homogeneous system of $n-1$ linear equations on $n$ partial derivatives of the function $U$ which has infinitely many solutions.

It is straightforward to verify that owing to (\ref{even}) the following relations
\bea
\frac{1}{\sum\limits_{k=0}^{n-1}\alpha_k\sigma_{0,k}^{n}\rho_k}\frac{\partial U}{\partial x_i}=
\frac{1}{\sum\limits_{k=0}^{n-1}\alpha_k\sigma_{1,k}^{n}\rho_k}\frac{\partial U}{\partial x_i^{(2)}}=...
=\frac{1}{\sum\limits_{k=0}^{n-1}\alpha_k\sigma_{n-1,k}^{n}\rho_k}\frac{\partial U}{\partial x_i^{2(n-1)}}
\nonumber
\eea
hold. When verifying these formulae, the identities (\ref{i}) prove to be helpful. Thus, we have the following ansatz for the function $U$
\bea\label{INT}
U=U\left(\sum_{k,m=0}^{n-1}\alpha_k\sigma_{m,k}^{n}\rho_k x_i^{(2m)}\right)=
U\left(\sum_{k=0}^{n-1}sign(\alpha_k)\sqrt{|\alpha_k|\rho_k}x_i^{k}\right)
\eea
which preserves the Hamiltonian structure of the equations
\bea
\{x_i^{(k)},\mathcal{H}_{int}\}=x_i^{(k+1)},\qquad k=0,1,..,2n-2.
\nonumber
\eea
Then the Hamiltonian $\mathcal{H}_{int}$ can be viewed as describing a deformed PU oscillator whose equations of motion read
\bea\label{PU1}
\sum_{k=0}^{n}\sigma_{k}^{n}x_i^{(2k)}-\{x_i^{(2n-1)},U\}=0.
\eea
In a recent work \cite{KL} (see also \cite{KLS}), this modification of the one-dimensional PU oscillator\footnote{About deformations of the PU oscillator see also \cite{Pavsic,Pavsic1} and references therein.} has been investigated with the use of the concept of the Lagrange anchor \cite{anchor1}. It was demonstrated that (\ref{PU1}) follows from the variational problem only if $\alpha_k=(-1)^{k+1}$. Otherwise, this equation can be viewed as a non-variational deformation of the original PU oscillator.
We thus conclude that the model in \cite{KL} can be treated as the unique generalization of the PU oscillator which is compatible with the Poisson structure (\ref{PB}).

The existence of the class of functions (\ref{INT}) allows one to construct new interacting nonvariational systems which admit canonical formulation with positive-definite Hamiltonian. It suffices to choose all $\alpha_k$ to be positive. Then the Hamiltonian
\bea\label{inter}
\tilde{\mathcal{H}}_{int}=\frac{1}{2}\sum_{k=0}^{n-1}(p_i^k p_i^k+\omega_k^2 x_i^k x_i^k)+\frac{1}{2}p_i p_i +U_1\left(\sum_{k=0}^{n-1}\sqrt{\alpha_k\rho_k}x_i^{k}\right)U_2(y_i)
\eea
together with the Poisson structure which results from (\ref{PB}) and the relations
\bea
\{y_i,p_j\}=\d_{ij},\qquad p_i=\dot{y}_i,
\nonumber
\eea
correspond to the PU oscillator and nonrelativistic particle which interact with each other.
The dynamics of this model obeys the following system of equations
\bea\label{EOMFP}
\sum_{k=0}^{n}\sigma_{k}^{n}x_i^{(2k)}-U_2\{x_i^{(2n-1)},U_1\}=0,\qquad \ddot{y}_i-U_1\frac{\partial U_2}{\partial y_i}=0.
\eea
If the function $U=U_1 U_2$ is positive-definite, then the Hamiltonian (\ref{inter}) also has the same property.

In a similar fashion one can realize the coupling of the PU oscillator with the harmonic oscillator, with another PU oscillator, etc. Many particle generalization is straightforward as well.

\vskip 0.5cm
\noindent
{\bf{\large{3. $\mathcal{N}=2$ supersymmetric PU oscillator}}}
\vskip 0.5cm
\noindent
{\bf{3.1. An alternative Hamiltonian formalism}}
\vskip 0.5cm

Let us generalize the results obtained in the previous section to the case of $\mathcal{N}=2$ supersymmetric PU oscillator \cite{Masterov,Masterov_1}. Apart from $x_i$ this model is described by extra bosonic coordinates $z_i$ and by fermionic coordinates $\psi_i$, $\bar{\psi}_i$ which are complex conjugate to each other. The evolution of $x_i$ is governed by (\ref{EOM}) while the dynamics of other variables is described by
\bea\label{EOM1}
\prod_{m=1-n}^{n-1}\left(\frac{d}{dt}-i\omega_m\right)\psi_i=0,\; \prod_{m=1}^{n-1}\left(\frac{d^2}{dt^2}+\omega_m^{2}\right)z_i=0, \; \prod_{m=1-n}^{n-1}\left(\frac{d}{dt}+i\omega_m\right)\bar{\psi}_i=0,
\eea
where we denoted $\omega_{-k}=-\omega_{k}$.

The equations (\ref{EOM}), (\ref{EOM1}) can be derived from the action functional \cite{Masterov_1}
\bea\label{act1}
\begin{aligned}
&
S=\frac{1}{2}\int\,dt\left(x_i\prod_{k=0}^{n-1}\left(\frac{d^2}{dt^2}+\omega_k^{2}\right)x_i-
i\psi_i\prod_{k=1-n}^{n-1}\left(\frac{d}{dt}+i\omega_k\right)\bar{\psi}_i-\right.
\\[3pt]
&
\left.\qquad\qquad\;\,-i\bar{\psi}_i\prod_{k=1-n}^{n-1}\left(\frac{d}{dt}-i\omega_k\right)\psi_i-
z_i\prod_{k=1}^{n-1}\left(\frac{d^2}{dt^2}+\omega_k^{2}\right)z_i\right).
\end{aligned}
\eea
This action is invariant under time translations. The corresponding Noether integral of motion can be presented in the form \cite{Masterov_1}
\bea\label{HN=2}
H=\sum_{k=0}^{n-1}(-1)^{k+1}J_k+\sum_{k=-n+1}^{n-1}(-1)^{k+1}F_k+\sum_{k=1}^{n-1}(-1)^{k+1}J_{-k},
\eea
where $J_k$ with $k=0,1,..,n-1$ are defined in (\ref{J}), while\footnote{For the fourth-order $\,\mathcal{N}=2$ supersymmetric PU oscillator the integral of motion $J_{-1}=\frac{1}{2}\left(\frac{d z_i}{dt}\right)^2+\frac{1}{2}\omega_1^2 z_i^2$ corresponds to the symmetry transformations $\d z_i=-\frac{d z_i}{dt}\epsilon_{-1}$, $\{z_i,\dot{z}_j\}=\d_{ij}$.}
\bea\label{JFz}
\begin{aligned}
&
F_k=\omega_k\beta_k\prod_{m=-n+1\atop m\neq k}^{n-1}\left(\frac{d}{dt}-i\omega_m\right)\psi_i \prod_{m=-n+1\atop m\neq k}^{n-1}\left(\frac{d}{dt}+i\omega_m\right)\bar{\psi}_i,\qquad
\beta_k=\frac{(-1)^{k+n-1}}{\prod\limits_{i=-n+1\atop i\neq k}^{n-1}(\omega_{i}-\omega_{k})},
\\[5pt]
&
J_{-k}=\frac{\mu_{k}}{2}\left(\prod_{m=1\atop m\neq k}^{n-1}\left(\frac{d^2}{dt^2}+\omega_m^2\right)\frac{d z_i}{dt}\right)^{2}+\frac{\mu_{k}\omega_k^2}{2}\left(\prod_{m=1\atop m\neq k}^{n-1}\left(\frac{d^2}{dt^2}+\omega_m^2\right)z_i\right)^2,\; \mu_{k}=\frac{(-1)^{k+1}}{\prod\limits_{i=1\atop i\neq k}^{n-1}(\omega_{i}^{2}-\omega_{k}^{2})}.
\end{aligned}
\eea

The quantities $F_k$ and $J_{-k}$ are integrals of motion which correspond to the following symmetry transformations
\bea
&&
\d\psi_i=-\beta_k\prod_{m=-n+1\atop m\neq k}^{n-1}\left(\frac{d}{dt}-i\omega_m\right)\frac{d\psi_i}{dt}\varepsilon_k,\quad
\d\bar{\psi}_i=-\beta_k\prod_{m=-n+1\atop m\neq k}^{n-1}\left(\frac{d}{dt}+i\omega_m\right)\frac{d\bar{\psi}_i}{dt}\varepsilon_k,
\nonumber
\\[5pt]
&&
\qquad\qquad\qquad\qquad\qquad\d z_i=\mu_{k}\prod_{m=1\atop m\neq k}^{n-1}\left(\frac{d^2}{dt^2}+\omega_m^{2}\right)\frac{d z_i}{dt}\,\epsilon_{-k},
\nonumber
\eea
where $\varepsilon_k$ and $\epsilon_{-k}$ are infinitesimal parameters of the transformations corresponding to $F_k$ and $J_{-k}$, respectively.

If we deform the Hamiltonian (\ref{HN=2}) as follows\footnote{As usual, all coefficients are arbitrary nonzero constants.}
\bea\label{H+}
\mathcal{H}=\sum_{k=-n+1}^{n-1}\left(\alpha_k J_k+\gamma_k F_k\right)
\eea
and require that the equations (\ref{canon2}) and their analogues for the variables $\psi_i$, $\bar{\psi}_i$, $z_i$ are satisfied, then the following graded Poisson bracket is found
\bea\label{PSN=2}
\begin{aligned}
&
\{A,B\}=\sum_{s,m=0}^{2n-1}w_{sm}\frac{\partial A}{\partial x_i^{(s)}}\frac{\partial B}{\partial x_i^{(m)}}+
\sum_{s,m=0}^{2n-3}\tilde{w}_{sm}\frac{\partial A}{\partial z_i^{(s)}}\frac{\partial B}{\partial z_i^{(m)}}+
\\[5pt]
&
\qquad\quad\;+\sum_{s,m=0}^{2n-2}f_{sm}\left(\frac{\overleftarrow{\partial}A}{\partial\psi_i^{(s)}}\frac{\overrightarrow{\partial}B}{\partial{\bar{\psi}}_i^{(m)}}+ \frac{\overleftarrow{\partial}A}{\partial\bar{\psi}_i^{(s)}}\frac{\overrightarrow{\partial}B}{\partial{\psi}_i^{(m)}}\right),
\end{aligned}
\eea
where the coefficients $f_{sm}$ and $\tilde{w}_{sm}$ are defined by
\bea\label{PBN=2}
f_{sm}=i^{s-m-1}\sum_{k=-n+1}^{n-1}\frac{\omega_k^{s+m}\beta_k}{\gamma_k},
\quad
\tilde{w}_{sm}=\left\{\begin{aligned}
&
0,&& \mbox{$s+m$ - even};\\[5pt]
&
(-1)^\frac{s+m-(-1)^{s}}{2}\sum_{k=1}^{n-1}\frac{\omega_k^{s+m-1}\mu_{k}}{\alpha_{-k}},&& \mbox{$s+m$ - odd},
\end{aligned}\right.
\eea
and $w_{sm}$ was introduced in (\ref{PB}).
The fact that this bracket produces the Hamiltonian equations for all the variables of the configuration space of an $\,\mathcal{N}=2$ supersymmetric PU oscillator can be verified in the same way as it has been done for the Poisson structure (\ref{PB}) in the previous section (some helpful formulae are given in Appendix).

Let us introduce the oscillator coordinates \cite{Pais,Masterov_1}
\bea\label{varN=2}
\begin{aligned}
&
x_i^{-k}=\sqrt{|\alpha_{-k}|\tilde{\rho}_{k}}\prod_{m=1\atop m\neq k}^{n-1}\left(\frac{d^2}{dt^2}+\omega_m^2\right)z_i,\quad p_i^{-k}=sign(\alpha_{-k})\frac{d x_i^{-k}}{dt},\quad k=1,2,..,n-1,
\\[5pt]
&
\psi_i^{k}=\sqrt{|\gamma_k|\beta_k}\prod_{m=-n+1\atop m\neq k}^{n-1}\left(\frac{d}{dt}-i\omega_m\right)\psi_i, \quad \bar{\psi}_i^{k}=(\psi_i^k)^{*},\quad
k=-n+1,..,n-1,
\end{aligned}
\eea
where $(\cdot)^{*}$ stands for the complex conjugation. These coordinates obey the structure relations
\bea\label{canN=2}
\begin{aligned}
&
\{x_i^{-k},x_j^{-m}\}=\d_{ij}\d_{km},\qquad
\{\psi_i^{k},\bar{\psi}_j^{m}\}=-i\,sign(\gamma_k)\d_{ij}\d_{km}.
\end{aligned}
\eea
In terms of coordinates (\ref{var}), (\ref{varN=2}) the Hamiltonian (\ref{H+}) takes the form
\bea
\mathcal{H}=\frac{1}{2}\sum_{k=-n+1}^{n-1}\left(sign(\alpha_{k})(p_i^k p_i^k+\omega_k^2 x_i^k x_i^k)+2sign(\gamma_k)\omega_k\psi_i^k\bar{\psi}_i^k\right).
\nonumber
\eea
As in the bosonic case, one has a Hamiltonian which is more suitable for quantization only if all coefficients $\alpha_k$ are positive.
On the other hand, the presence of the negative coefficients $\gamma_k$ leads to the structure relations $\{\psi_i^{k},\bar{\psi}_j^{m}\}=i\d_{ij}\d_{km}$ (see (\ref{canN=2})).
As was demonstrated in \cite{Masterov_1}, on quantization these relations immediately bring about the negative norm states.
So, one has to set all the coefficients $\gamma_k$ to be positive as well.
The corresponding Hamiltonian reads
\bea\label{Hosc}
\mathcal{H}=\frac{1}{2}\sum_{k=-n+1}^{n-1}(p_i^k p_i^k+\omega_k^2 x_i^k x_i^k+2\omega_k\psi_i^k\bar{\psi}_i^k).
\eea

Along with the Hamiltonian, the full formulation of an $\,\mathcal{N}=2$ supersymmetric PU oscillator involves supercharges. We obtain these in the next subsection.

\vskip 0.5cm
\noindent
{\bf{3.2. Supercharges and other integrals of motion}}
\vskip 0.5cm

The form of the Hamiltonian (\ref{Hosc}) allows one to use the action functional
\bea\label{action}
S=\frac{1}{2}\int dt\sum_{k=-n+1}^{n-1}\left(\dot{x}_i^{k}\dot{x}_i^k-\omega_k^2 x_i^k x_i^k+i\psi_i^k\dot{\bar{\psi}}_i^k+i\bar{\psi}_i^k\dot{\psi}_i^k-2\omega_k\psi_i^k\bar{\psi}_i^k\right)
\eea
so as to obtain integrals of motion for the original $\,\mathcal{N}=2$ supersymmetric PU oscillator. In particular, the Hamiltonian (\ref{Hosc}) is the Noether integral of motion which corresponds to the invariance under time translations. Moreover, a conventional Hamiltonian formulation for the model (\ref{action}) leads to the structure relations (\ref{canon}), (\ref{canN=2}) (for technical details see, e.g. \cite{Masterov_1}). The supersymmetry transformations
\bea\label{super}
\d x_i^{k}=\psi_i^{k}\alpha+\bar{\psi}_i^{k}\bar{\alpha}, \qquad \d\psi_i^{k}=-(i\dot{x}_i^{k}-\omega_k x_i^{k})\bar{\alpha}, \qquad \d\bar{\psi}_i^{k}=-(i\dot{x}_i^{k}+\omega_k x_i^{k})\alpha,
\eea
lead to the desirable supercharges
\bea\label{Qosc}
&&
Q=\sum_{k=-n+1}^{n-1}\psi_i^k\left(p_i^k-i\omega_k x_i^k\right),
\qquad
\bar{Q}=\sum_{k=-n+1}^{n-1}\bar{\psi}_i^k\left(p_i^k+i\omega_k x_i^k\right).
\eea

The action functional (\ref{action}) is also invariant under the bosonic and fermionic translations
\bea
\d x_i^k=\cos{(\omega_k t)}a_i^k+\frac{1}{\omega_k}\sin{(\omega_k t)}b_i^k,\quad \d\psi_i^k=e^{it\omega_k}\alpha_i^k,\quad \d\bar{\psi}_i^k=e^{-it\omega_k}\bar{\alpha}_i^k,
\nonumber
\eea
as well as under rotations
\bea
\d x_i^k=\omega_{ij}x_j^k,\quad \d\psi_i^k=\omega_{ij}\psi_j^k,\quad \d\bar{\psi}_i^k=\omega_{ij}\bar{\psi}_j^k,\qquad\mbox{where}\quad\omega_{ij}=-\omega_{ji},
\nonumber
\eea
and the $U(1)$ $R$-symmetry transformations
\bea
\d\psi_i^k=i\nu\psi_i^k,\qquad \d\bar{\psi}_i^k=-i\nu\bar{\psi}_i^k.
\nonumber
\eea
The Noether theorem yields the integrals of motion
\bea\label{intN=2}
\begin{aligned}
&
P_i^k=\cos{(\omega_k t)}p_i^k+\omega_k\sin{(\omega_k t)}x_i^k,\quad\; \Psi_i^{k}=e^{-i\omega_k t}\psi_i^k,\quad M_{ij}=\sum_{k=-n+1}^{n-1}-x^k_{[i}p^k_{j]}+i\psi^k_{[i}\bar{\psi}^k_{j]},
\\[5pt]
&
X_i^{k}=\frac{1}{\omega_k}\sin{(\omega_k t)}p_i^k-\cos{(\omega_k t)}x_i^k,\quad \bar{\Psi}_i^{k}=e^{it\omega_k}\bar{\psi}_i^k,\quad\;\, J=\sum_{k=-n+1}^{n-1}\psi_i^{k}\bar{\psi}_i^k.
\end{aligned}
\eea
which obey
\bea
\begin{aligned}
&
\{Q,\bar{Q}\}=-2iH, && \{M_{ij},A_s^k\}=A_i^s\d_{jk}-A_j^s\d_{ik},&& A_i^k=P_i^k,X_i^k,\Psi_i^k,\bar{\Psi}_i^k
\\[5pt]
&
\{H,\bar{\Psi}_i^k\}=i\omega_k\bar{\Psi}_i^k,&& \{H,\Psi_i^k\}=-i\omega_k\Psi_i^k,&& \{H,P_i^k\}=-\omega_k^2 X_i^k, && \{H,X_i^k\}=P_i^k,
\\[5pt]
&
\{J,Q\}=-i Q,&& \{Q,\bar{\Psi}_i^k\}=-i P_i^k+\omega_k X_i^k, &&\{Q,P_i^k\}=-i\omega_k\Psi_i^k, && \{Q,X_i^k\}=\Psi_i^k,
\\[5pt]
&
\{J,\bar{Q}\}=i\bar{Q},&& \{\bar{Q},\Psi_i^k\}=-i P_i^k-\omega_k X_i^k, &&\{\bar{Q},P_i^k\}=i\omega_k\bar{\Psi}_i^k, && \{\bar{Q},X_i^k\}=\bar{\Psi}_i^k,
\\[5pt]
&
\{\Psi_i^k,\bar{\Psi}_j^m\}=-i\d_{ij}\d_{km},&& \{P_i^k,X_j^m\}=\d_{km}\d_{ij}, &&\{J,\Psi_i^k\}=-i\Psi_i^k,&& \{J,\bar{\Psi}_i^k\}=i\bar{\Psi}_i^k.
\end{aligned}
\nonumber
\eea

Despite the fact that our alternative Hamiltonian formulation has been obtained beyond the conventional technique, we are still able to use the Noether theorem to construct integrals of motion of an $\,\mathcal{N}=2$ supersymmetric PU oscillator by using of the action functional (\ref{action}).

To conclude this section, let us clarify some details regarding quantization of an $\,\mathcal{N}=2$ supersymmetric PU oscillator with the Hamiltonian (\ref{Hosc}).
In order to construct a quantum counterpart of this model, we introduce the hermitian operators $\hat{x}_i^{k}$, $\hat{p}_i^{k}$ and the operators $\hat{\psi}_i^{k}$, $\hat{\bar{\psi}}_i^k$ which are hermitian conjugates of each other.
In accord with (\ref{canon}), (\ref{canN=2}), one finds the following nonvanishing (anti)commutation relations
\bea
[\hat{x}_i^k,\hat{p}_j^m]=i\hbar\d_{km}\d_{ij},\qquad \{\hat{\psi}_i^k,\hat{\bar{\psi}}_j^m\}=\hbar\d_{km}\d_{ij}.
\nonumber
\eea
Let us also introduce the creation $\bar{a}_i^{k}$, $\bar{c}_i^{k}$ and annihilation $a_i^k=\left(\bar{a}_i^k\right)^\dagger$, $c_i^k=\left(\bar{c}_i^k\right)^\dagger$ operators
\bea
&&
a_i^k=\sqrt{\frac{\omega_{|k|}}{2\hbar}}\hat{x}_i^k+\frac{i}{\sqrt{2\omega_{|k|}\hbar}}\hat{p}_i^k,\qquad \bar{a}_i^k=\sqrt{\frac{\omega_{|k|}}{2\hbar}}\hat{x}_i^k-\frac{i}{\sqrt{2\omega_{|k|}\hbar}}\hat{p}_i^k,\quad\, k=-n+1,..,n-1,
\nonumber
\\[5pt]
&&
\qquad\quad c_i^{k}=\left\{
\begin{aligned}
&
\frac{1}{\sqrt{\hbar}}\hat{\bar{\psi}}_i^k,
\\[5pt]
&
\frac{1}{\sqrt{\hbar}}\hat{\psi}_i^k,
\end{aligned}\qquad\qquad\qquad\qquad
\right.
\bar{c}_i^{k}=\left\{
\begin{aligned}
&
\frac{1}{\sqrt{\hbar}}\hat{\psi}_i^k,\qquad\qquad k=0,1,..,n-1,
\\[5pt]
&
\frac{1}{\sqrt{\hbar}}\hat{\bar{\psi}}_i^k,\qquad\qquad k=-n+1,..,-1,
\end{aligned}
\right.
\nonumber
\eea
which obey the structure relations
\bea
[a_i^k,\bar{a}_j^m]=\d_{km}\d_{ij},\qquad \{c_i^k,\bar{c}_j^m\}=\d_{km}\d_{ij}.
\nonumber
\eea

The quantum Hamiltonian can be written in the form\footnote{We choose the Weyl ordering for the fermions  $\psi_i^{k}\bar{\psi}_i^{k}\;\rightarrow\;\frac{1}{2}\left(\hat{\psi}_i^{k}\hat{\bar{\psi}}_i^{k}- \hat{\bar{\psi}}_i^{k}\hat{\psi}_i^{k}\right)$.}
\bea
\hat{H}=\sum_{k=-n+1}^{n-1}\hbar\omega_{|k|}\left(\bar{a}_i^{k}a_i^k+\bar{c}_i^{k}c_i^k\right).
\eea
Evidently, this variant of the quantum $\,\mathcal{N}=2$ supersymmetric PU oscillator has a stable ground state as well as a bounded from below energy spectrum. The Hilbert space does not contain negative norm states.

\vskip 0.5cm
\noindent
{\bf {\large 4. Conclusion}}
\vskip 0.5cm
\noindent

In this paper we have constructed an alternative canonical formulation for the PU oscillator of order $2n$ with positive-definite Hamiltonian following the method originally proposed in \cite{Kosinski}.
The corresponding Poisson structure has been used in order to find possible generalizations of the PU oscillator which are compatible with the alternative Hamiltonian formulation.
A procedure to construct interacting many-body mechanics whose dynamics is governed by a system of nonvariational equations has been proposed.
An alternative Hamiltonian formulation for an $\,\mathcal{N}=2$ supersymmetric PU oscillator has been constructed as well.

In Refs. \cite{Galajinsky_1,PU} (see also \cite{Andrzejewski}-\cite{Masterov_4}) it was shown that the PU oscillator is conformal invariant provided frequencies of oscillation form the arithmetic sequence $\omega_k=(2k+1)\omega_0$.
It is interesting to investigate how the conformal invariance can be realized within the framework of the alternative Hamiltonian formulation.
Possible generalizations of an $\,\mathcal{N}=2$ supersymmetric PU oscillator which are compatible with the graded Poisson structure are worth studying as well.
This requires more a sophisticated construction because one has to deform both the Hamiltonian and the supercharges in such a way that these deformations preserve both the Hamiltonian structure of the equations of motions and the algebra.

In Refs. \cite{KLS,KL} the stability of the PU oscillator has been investigated with the aid of the concept of the Lagrange anchor \cite{anchor1}. The latter can be applied to study higher-derivative field theories as well \cite{KLS}. It is of interest to see how this method works for an $\,\mathcal{N}=2$ supersymmetric PU oscillator.

\vskip 0.8cm
\noindent
{\bf{\large{Acknowledgements}}}
\vskip 0.5cm

\noindent
We thank M. Pav\v{s}i\v{c} for the correspondence and D.S. Kaparulin for useful discussions. This work was supported by the Dynasty Foundation, the RFBR grant 14-02-31139-Mol, the MSE program "Nauka" under
the project 3.825.2014/K, and the TPU grant LRU.FTI.123.2014.

\newpage
\vskip 0.8cm
\noindent
{\bf{\large{Appendix. The proof of identities (\ref{i})}}}
\vskip 0.5cm
\noindent
{\bf a)}
Let us prove that
$$
\sum_{s=0}^{n-1}(-1)^s\omega_p^{2s}\sigma_{s,k}^{n}=\frac{(-1)^k}{\rho_k}\d_{kp}.\eqno{(A1)}
$$

First take into account that $\sigma_{s,k}^{n}$ can be represented in the form \cite{Ernst}
\bea
\sigma_{s,k}^{n}=\frac{1}{V\rho_k}\left|
\begin{aligned}
& 1 && \omega_0^2 &&  ... &&\omega_{0}^{2s-2} &&\omega_{0}^{2s+2} &&... && \omega_0^{2n-2}
\\[2pt]
& 1 && \omega_1^2 &&  ... &&\omega_{1}^{2s-2} &&\omega_{1}^{2s+2} &&... && \omega_1^{2n-2}
\\[2pt]
& ... && ... && ...  && ... &&... &&... && ...
\\[2pt]
& 1 && \omega_{k-1}^2 &&  ... &&\omega_{k-1}^{2s-2} &&\omega_{k-1}^{2s+2} &&... && \omega_{k-1}^{2n-2}
\\[2pt]
& 1 && \omega_{k+1}^2 &&  ... &&\omega_{k+1}^{2s-2} &&\omega_{k+1}^{2s+2} &&... && \omega_{k+1}^{2n-2}
\\[2pt]
& ... && ... && ...  && ... && ...&&... && ...
\\[2pt]
& 1 && \omega_{n-1}^2 && ... &&\omega_{n-1}^{2s-2} &&\omega_{n-1}^{2s+2} &&... && \omega_{n-1}^{2n-2}
\end{aligned}
\right|,
\nonumber
\eea
where $V=\prod\limits_{i_1<i_2=0}^{n-1}(\omega_{i_2}^2-\omega_{i_1}^2)$ is the Vandermonde determinant. Then the identity ($A1$) can be written in the form
\bea
\sum_{s=0}^{n-1}(-1)^s\omega_p^{2s}\sigma_{s,k}^{n}=\frac{1}{V\rho_k}
\left|
\begin{aligned}
& 1 && \omega_p^2 && \omega_p^4 && ...  && \omega_p^{2n-4} && \omega_p^{2n-2}
\\[2pt]
& 1 && \omega_0^2 && \omega_0^4 && ...  && \omega_0^{2n-4} && \omega_0^{2n-2}
\\[2pt]
& ... && ... && ... && ... && ...  && ...
\\[2pt]
& 1 && \omega_{k-1}^2 && \omega_{k-1}^4  && ..  && \omega_{k-1}^{2n-4} && \omega_{k-1}^{2n-2}
\\[2pt]
& 1 && \omega_{k+1}^2 && \omega_{k+1}^4  && .. && \omega_{k+1}^{2n-4} && \omega_{k+1}^{2n-2}
\\[2pt]
& ... && ... && ... && ...  && ... &&  ...
\\[2pt]
& 1 && \omega_{n-1}^2 && \omega_{n-1}^4 && ... && \omega_{n-1}^{2n-4} && \omega_{n-1}^{2n-2}.
\end{aligned}
\right|
\nonumber
\eea
If $p=k$ the determinant is equal to $(-1)^k V$. Otherwise, it is zero. So, one has
\bea
\sum_{s=0}^{n-1}(-1)^s\omega_p^{2s}\sigma_{s,k}^{n}=\frac{1}{V\rho_k}(-1)^k V\d_{kp}=\frac{(-1)^k}{\rho_k}\d_{kp}.
\nonumber
\eea
\vskip 0.5cm
\noindent
{\bf b)}
Let us prove another identity from (\ref{i})
$$
\sum_{k=0}^{n-1}(-1)^k(-\omega_k^{2})^{s}\sigma_{p,k}^n\rho_k=\delta_{sp},\;\mbox{where $s=0,1,..,n-1$.}\eqno{(A2)}
$$
At the first step, let us show that it holds for $p=0$. Taking into account that $\sigma_{0,k}^{n}=\prod\limits_{i=0\atop i\neq k}^{n-1}\omega_i^2$, one obtains
\bea
&&
\sum_{k=0}^{n-1}(-1)^k(-\omega_k^{2})^{s}\sigma_{0,k}^n\rho_k=
(-1)^{s}\prod_{i=0}^{n-1}\omega_i^2\sum_{k=0}^{n-1}(-1)^k\omega_k^{2(s-1)}\rho_k=
\nonumber
\\[5pt]
&&
=\frac{(-1)^{s}}{V}\prod\limits_{i=0}^{n-1}\omega_i^2\left|
\begin{aligned}
&\omega_0^{2(s-1)} && \omega_1^{2(s-1)} &&  ... && \omega_{n-1}^{2(s-1)}
\\[5pt]
& 1 && 1 && ... && 1
\\[5pt]
& \omega_0^2 && \omega_1^2 &&... && \omega_{n-1}^{2}
\\[5pt]
& ... && ... &&  ... && ...
\\[5pt]
&\omega_{0}^{2(n-2)} && \omega_{1}^{2(n-2)} &&  ... && \omega_{n-1}^{2(n-2)}
\end{aligned}
\right|,
\nonumber
\eea
which implies that
\bea
\sum_{k=0}^{n-1}(-1)^k(-\omega_k^{2})^{s}\sigma_{0,k}^n\rho_k\sim\d_{s,0}
\nonumber
\eea
for $0\leq s\leq n-1$.
Let us consider the case $p=s=0$ and prove by induction that
\bea
\sum_{k=0}^{n-1}(-1)^k\sigma_{0,k}^n\rho_k=1.
\nonumber
\eea
Indeed, it is straightforward to verify that this identity holds for $n=2$. Then the chain of identical transformations
\bea
\begin{aligned}
&
\sum_{k=0}^{n}(-1)^k\sigma_{0,k}^{n+1}\rho_k=(-1)^n\sigma_{0,n}^{n+1}\rho_n+\sum_{k=0}^{n-1}(-1)^k\sigma_{0,k}^{n+1}\rho_k=
(-1)^n\rho_n\prod_{i=0}^{n-1}\omega_i^2+\sum_{k=0}^{n-1}(-1)^k\omega_n^2\sigma_{0,k}^{n}\rho_k=
\\[5pt]
&
=\frac{\prod\limits_{i=0}^{n-1}\omega_i^2}{\prod\limits_{m=0}^{n-1}(\omega_m^2-\omega_n^2)}+\sum_{k=0}^{n-1}\frac{\omega_n^2\sigma_{0,k}^{n}}
{\prod\limits_{m=0\atop m\neq k}^{n}(\omega_m^2-\omega_k^2)}=
\frac{\prod\limits_{i=0}^{n-1}\omega_i^2}{\prod\limits_{m=0}^{n-1}(\omega_m^2-\omega_n^2)}+\sum_{k=0}^{n-1}\frac{(\omega_n^2-\omega_k^2+\omega_k^2)\sigma_{0,k}^{n}}{\prod\limits_{m=0\atop m\neq k}^{n-1}(\omega_m^2-\omega_k^2)(\omega_n^2-\omega_k^2)}=
\\[5pt]
&
=\frac{\prod\limits_{i=0}^{n-1}\omega_i^2}{\prod\limits_{m=0}^{n-1}(\omega_m^2-\omega_n^2)}+\sum_{k=0}^{n-1}\frac{\sigma_{0,k}^{n}}{\prod\limits_{m=0\atop m\neq k}^{n-1}(\omega_m^2-\omega_k^2)}+\sum_{k=0}^{n-1}\frac{\omega_k^2\sigma_{0,k}^{n}}{\prod\limits_{m=0\atop m\neq k}^{n}(\omega_m^2-\omega_k^2)}=
\\[5pt]
&
=\sum_{k=0}^{n-1}(-1)^k\sigma_{0,k}^n\rho_k+\frac{\prod\limits_{i=0}^{n-1}\omega_i^2}{\prod\limits_{m=0}^{n-1}(\omega_m^2-\omega_n^2)}+
\frac{1}{\omega_n^2}\sum_{k=0}^{n-1}\frac{\omega_k^2\sigma_{0,k}^{n+1}}{\prod\limits_{m=0\atop m\neq k}^{n}(\omega_m^2-\omega_k^2)}=
\end{aligned}
\nonumber
\eea
\bea
\begin{aligned}
&
=\sum_{k=0}^{n-1}(-1)^k\sigma_{0,k}^n\rho_k+\frac{\prod\limits_{i=0}^{n-1}\omega_i^2}{\prod\limits_{m=0}^{n-1}(\omega_m^2-\omega_n^2)}+
\frac{1}{\omega_n^2}\left(\sum_{k=0}^{n}\frac{\omega_k^2\sigma_{0,k}^{n+1}}{\prod\limits_{m=0\atop m\neq k}^{n}(\omega_m^2-\omega_k^2)}-
\frac{\prod\limits_{i=0}^n\omega_i^2}{\prod\limits_{m=0}^{n-1}(\omega_m^2-\omega_n^2)}\right)=
\\[5pt]
&
\sum_{k=0}^{n-1}(-1)^k\sigma_{0,k}^n\rho_k+\frac{1}{\omega_n^2}\sum_{k=0}^{n}(-1)^k\omega_k^2\sigma_{0,k}^{n+1}\rho_k=\sum_{k=0}^{n-1}(-1)^k\sigma_{0,k}^n\rho_k=1
\end{aligned}
\nonumber
\eea
results in
$$
\sum_{k=0}^{n-1}(-1)^k(-\omega_k^{2})^{s}\sigma_{0,k}^n\rho_k=\d_{s,0}.\eqno{(A3)}
$$

At the next step, let us prove that the identity ($A2$) holds for $s\geq p$. To this end, we use the following relation
\bea
\sigma_{p,k}^{n}=\Delta_{p,k}^{n}\sigma_{0,k}^n, \quad\mbox{where}\qquad \Delta_{p,k}^{n}=\sum_{i_1<i_2<..<i_p=0\atop i_1,i_2,..,i_p\neq k}^{n-1}\frac{1}{\omega_{i_1}^2\omega_{i_2}^2..\omega_{i_p}^2},\quad\Delta_{0,k}^{n}=1.
\nonumber
\eea
Let us also define
\bea
\Delta_{p}^{n}=\sum_{i_1<i_2<..<i_p=0}^{n-1}\frac{1}{\omega_{i_1}^2\omega_{i_2}^2..\omega_{i_p}^2}.
\nonumber
\eea
Then it is easy to show that
\bea
\Delta_{p,k}^{n}=\Delta_{p}^{n}-\frac{1}{\omega_k^2}\Delta_{p-1}^{n}\; \rightarrow\; \Delta_{p,k}^{n}=\sum_{m=0}^{p}\frac{(-1)^m}{\omega_k^{2m}}\Delta_{p-m}^{n},
\nonumber
\eea
where, by definition, we have set $\Delta_{0}^{n}=1$.
At this stage one finds
\bea
&&
\sum_{k=0}^{n-1}(-1)^k(-\omega_k^{2})^{s}\sigma_{p,k}^n\rho_k=
\sum_{k=0}^{n-1}(-1)^k(-\omega_k^{2})^{s}\rho_k
\sum_{m=0}^{p}\frac{(-1)^m}{\omega_k^{2m}}\Delta_{p-m}^{n}\sigma_{0,k}^{n}=
\nonumber
\\[5pt]
&&
=\sum_{m=0}^{p}\Delta_{p-m}^{n}\sum_{k=0}^{n-1}(-1)^{k}(-\omega_k^2)^{s-m}\sigma_{0,k}^{n}\rho_k.
\nonumber
\eea
According to ($A3$) the latter expression is equal to zero if $s>p$. If $s=p$, one has
\bea
\sum_{m=0}^{p}\Delta_{p-m}^{n}\sum_{k=0}^{n-1}(-1)^{k}(-\omega_k^2)^{p-m}\sigma_{0,k}^{n}\rho_k=\sum_{m=0}^{p}\Delta_{p-m}^{n}\d_{pm}=\Delta_{0}^n=1.
\nonumber
\eea
So, we have shown that the identity ($A2$) holds for $n-1\geq s\geq p$. In order to finish the proof, one needs to use the relation
\bea
\sigma_{p,k}^{n}=\sigma_{p+1}^n-\omega_k^2\sigma_{p+1,k}^n\;\rightarrow\; \sigma_{p,k}^{n}=\sum_{r=0}^{n-p-1}(-1)^{r}\omega_{k}^{2r}\sigma_{p+r+1}^{n},
\nonumber
\eea
where we have taken into account that $\sigma_{n-1,k}^n=\sigma_{n}^{n}=1$.
Then one finds
\bea
\begin{aligned}
&
\sum_{k=0}^{n-1}(-1)^k(-\omega_k^{2})^{s}\sigma_{p,k}^n\rho_k=
\sum_{k=0}^{n-1}(-1)^k(-\omega_k^{2})^{s}\rho_k\sum_{r=0}^{n-p-1}(-1)^{r}\omega_{k}^{2r}\sigma_{p+r+1}^{n}=
\\[5pt]
&
=\sum_{r=0}^{n-p-1}\sigma_{p+r+1}^{n}\sum_{k=0}^{n-1}(-1)^k(-\omega_k^2)^{r+s}\rho_k=\sum_{r=0}^{n-p-1}(-1)^{r+s}\frac{\sigma_{p+r+1}^{n}}{V}
\left|
\begin{aligned}
&\omega_0^{2(r+s)} && \omega_1^{2(r+s)} && ... && \omega_{n-1}^{2(r+s)}
\\[5pt]
& 1 && 1 &&  ... && 1
\\[5pt]
& \omega_0^2 && \omega_1^2 && ... && \omega_{n-1}^{2}
\\[5pt]
& ... && ... && ... && ...
\\[5pt]
&\omega_{0}^{2(n-2)} && \omega_{1}^{2(n-2)} && ... && \omega_{n-1}^{2(n-2)}
\end{aligned}
\right|.
\end{aligned}
\nonumber
\eea
So, if the inequalities $0\leq r+s\leq n-2$ are satisfied, this expression is equal to zero. This is true for $s<p$. Thus, the identity ($A2$) is proved.

\vskip 0.5cm
\noindent
{\bf c)}
In order to show that the identity
$$
\sum_{k=0}^{n-1}(-1)^k(-\omega_k^{2})^{n}\sigma_{p,k}^n\rho_k=-\sigma_p^{n}\eqno{(A4)}
$$
holds, let us make the following identical transformations
\bea
&&
\sum_{k=0}^{n-1}(-1)^k(-\omega_k^{2})^{n}\sigma_{p,k}^n\rho_k=(-1)^n\sum_{k=0}^{n-1}(-1)^{k}\omega_k^{2n}\Delta_{p,k}^{n}\sigma_{0,k}^{n}\rho_k=
\nonumber
\\[5pt]
&&
=(-1)^n\sum_{k=0}^{n-1}(-1)^{k}\omega_k^{2n}\sigma_{0,k}^{n}\rho_k\sum_{m=0}^{p}\frac{(-1)^m}{\omega_k^{2m}}\Delta_{p-m}^n=
\nonumber
\\[5pt]
&&
=(-1)^n\sum_{m=0}^{p}(-1)^m\Delta_{p-m}^n\prod_{r=0}^{n-1}\omega_r^2\sum_{k=0}^{n-1}(-1)^{k}\omega_k^{2(n-m-1)}\rho_k=
\nonumber
\\[5pt]
&&
=(-1)^n\sum_{m=0}^{p}(-1)^m\Delta_{p-m}^n\sigma_{0}^{n}\frac{1}{V}\sum_{k=0}^{n-1}(-1)^{k}\omega_k^{2(n-m-1)}
\prod_{i_1<i_2=0\atop i_1,i_2\neq k}^{n-1}(\omega_{i_2}^2-\omega_{i_1}^2)=
\nonumber
\eea
\bea
&&
=(-1)^n\sum_{m=0}^{p}(-1)^m\Delta_{p-m}^n\sigma_{0}^{n}\frac{1}{V}\left|
\begin{aligned}
&
\omega_0^{2(n-m-1)} && \omega_1^{2(n-m-1)} && ... && \omega_{n-1}^{2(n-m-1)}
\\[5pt]
&
1 && 1 && ... && 1
\\[5pt]
&
\omega_0^2 && \omega_1^2 && ... && \omega_{n-1}^2
\\[5pt]
&
... && ... && ... && ...
\\[5pt]
&
\omega_0^{2(n-2)} && \omega_1^{2(n-2)} && ... && \omega_{n-1}^{2(n-2)}
\end{aligned}
\right|=
\nonumber
\eea
\bea
=(-1)^n\sum_{m=0}^{p}(-1)^m\sigma_{p-m}^{n}\frac{1}{V}(-1)^{n-1}V\d_{m,0}=-\sigma_{p}^{n}.
\nonumber
\eea
which establishes ($A4$).

\vskip 0.5cm
\noindent
{\bf d)} When verifying the fact that the graded Poisson bracket (\ref{PSN=2}) produces the Hamiltonian equations of motion for all the variables of the configuration space of an $\,\mathcal{N}=2$ supersymmetric PU oscillator, the following analogues of the identities (\ref{i})
\bea
&&
\sum_{r=0}^{2n-2}(-1)^r\omega_q^r\tilde{\sigma}_{r,k}^{n}=\frac{(-1)^{n+k-1}}{\beta_k}\d_{qk}, \qquad
\sum_{r=0}^{n-2}(-1)^r\omega_q^{2r}\overline{\sigma}_{r,k}^{n}=\frac{(-1)^{k+1}}{\mu_{k}}\d_{qk},
\nonumber
\\[5pt]
&&
\sum_{k=-n+1}^{n-1}(-1)^{n+k-1}(-\omega_k)^s\tilde{\sigma}_{p,k}^{n}\beta_k=\left\{
\begin{aligned}
&
\d_{sp}, && s=0,1,..,2n-2;
\\[5pt]
&
-\tilde{\sigma}_{p}^{n}, && s=2n-1
\end{aligned}
\right.
\nonumber
\\[5pt]
&&
\sum_{k=1}^{n-1}(-1)^{k+1}(-\omega_k^2)^s\overline{\sigma}_{p,k}^{n}\mu_{k}=\left\{
\begin{aligned}
&
\d_{sp}, && s=0,1,..,n-2;
\\[5pt]
&
-\overline{\sigma}_{p}^{n}, && s=n-1
\end{aligned}
\right.
\nonumber
\eea
prove to be helpful. Here we denoted
\bea
&&
\tilde{\sigma}_{p,k}^{n}=\sum_{i_1<i_2<..<i_{2n-p-2}=-n+1\atop i_1,i_2,..,i_{2n-p-2}\neq k}^{n-1}\omega_{i_1}\omega_{i_2}..\omega_{i_{2n-p-2}},\qquad
\tilde{\sigma}_{p}^{n}=\sum_{i_1<i_2<..<i_{2n-p-1}=-n+1}^{n-1}\omega_{i_1}\omega_{i_2}..\omega_{i_{2n-p-1}},
\nonumber
\\[5pt]
&&
\quad\overline{\sigma}_{p,k}^{n}=\sum_{i_1<i_2<..<i_{n-p-2}=1\atop i_1,i_2,..,i_{n-p-2}\neq k}^{n-1}\omega_{i_1}^2\omega_{i_2}^2..\omega_{i_{n-p-2}}^2,\qquad\qquad\;
\overline{\sigma}_{p}^{n}=\sum_{i_1<i_2<..<i_{n-p-1}=1}^{n-1}\omega_{i_1}^2\omega_{i_2}^2..\omega_{i_{n-p-1}}^2.
\nonumber
\eea
By definition, $\tilde{\sigma}_{2n-2,k}^{n}=\tilde{\sigma}_{2n-1}^{n}=\overline{\sigma}_{n-2,k}^{n}=\overline{\sigma}_{n-1}^{n}=1$.

\end{document}